\documentclass[%
reprint,
 superscriptaddress,
 amsmath,amssymb,
 aps,
pra,
]{revtex4-2}

\usepackage{graphicx}% Include figure files
\usepackage{dcolumn}% Align table columns on decimal point
\usepackage{bm}% bold math
\usepackage{color}
\usepackage{natbib}
\begin{document}

\preprint{APS/123-QED}

%\title{Strong Polarization Observed from non-Polarized Dielectronic Recombination in Highly Charged Ions}%Tong-san
\title{Polarization measurements of the $2s$--$2p_{3/2}$ VUV transition in N$^{4+}$\\
excited by electron collisions}
%\title{Unanticipated strong polarization of a $J$=1/2 to 1/2 transition\\
%arising from unexpectedly large quantum interference effect}% Force line breaks with \\

\author{Nobuyuki Nakamura}
\email{n\_nakamu@ils.uec.ac.jp}
\affiliation{%
Institute for Laser Science, The University of Electro-Communications, Tokyo 182-8585, Japan
}

\author{Ryohko Ishikawa}
\affiliation{%
National Astronomical Observatory of Japan, Tokyo 181-8588, Japan
}%

\author{Motoshi Goto}
\affiliation{National Institute for Fusion Science, Toki, Gifu 509-5292, Japan
}

\date{\today}% It is always \today, today,
    % but any date may be explicitly specified

%%%%%%%%%%%%%%%%%%%%%%%%%%%%%%%%%%%%%%%%%%%%%%%%%%%%%%%%%%%%%
%%%%%%%%%%%%%%%%%%%%%%%%%%%%%%%%%%%%%%%%%%%%%%%%%%%%%%%%%%%%%
\begin{abstract}
We present the linear polarization of the $2s$--$2p_{3/2}$ transition in Li-like N$^{4+}$ excited by electron collisions, measured at electron energies between 85 and 1000~eV.
The measured polarizations are compared with theoretical values calculated using the Flexible Atomic Code (FAC).
Although the overall trend that the degree of polarization decreases with increasing electron energy is reproduced, the theoretical values are systematically larger in magnitude than the experimental values.
After examining possible experimental sources of depolarization, we suggest that the discrepancy likely originates from an overestimation in the theoretical model.
Further theoretical and experimental investigations along the isoelectronic sequence are needed to resolve this discrepancy.
\end{abstract}
%%%%%%%%%%%%%%%%%%%%%%%%%%%%%%%%%%%%%%%%%%%%%%%%%%%%%%%%%%%%%
%%%%%%%%%%%%%%%%%%%%%%%%%%%%%%%%%%%%%%%%%%%%%%%%%%%%%%%%%%%%%

\maketitle

%\tableofcontents

\newpage

%%%%%%%%%%%%%%%%%%%%%%%%%%%%%%%%%%%%%%%%%%%%%%%%%%%%%%%%%%%%%
%%%%%%%%%%%%%%%%%%%%%%%%%%%%%%%%%%%%%%%%%%%%%%%%%%%%%%%%%%%%%
\section{\label{sec:introduction}Introduction}
%%%%%%%%%%%%%%%%%%%%%%%%%%%%%%%%%%%%%%%%%%%%%%%%%%%%%%%%%%%%%
%%%%%%%%%%%%%%%%%%%%%%%%%%%%%%%%%%%%%%%%%%%%%%%%%%%%%%%%%%%%%

The polarization of collisionally excited radiation from ions is important for anisotropy diagnostics of plasmas \cite{Henoux1,Fujimoto3,Goto2,Goto3,Ramaiya1,Baronova1,Dubau2,Kieffer1}.
For such applications, it is essential to understand the polarization mechanisms arising from elementary processes, such as collisions of ions with directional electrons.
The degree of polarization of the emitted radiation reflects the anisotropy of magnetic sublevel populations.
Because even subtle differences in the magnetic sublevel populations are clearly reflected in the polarization, experimental investigations of polarization provide a stringent test of theoretical models that explicitly account for the magnetic sublevel populations.
%In this sense, polarization serves as a sensitive probe of the underlying excitation and decay dynamics.
Therefore, systematic investigations of polarization not only complement conventional cross-section measurements but also offer deeper insight into the underlying excitation and decay dynamics, as well as into the validity and limitations of detailed theoretical treatments that include magnetic sublevel structure.
In particular, the polarization of radiation from highly charged ions is important for the anisotropy diagnostics of hot laboratory and astrophysical plasmas, and also for studying relativistic effects on the polarization mechanism \cite{Beiersdorfer6,Nakamura34,Shah1,Bostock1}.

Among various transitions in highly charged ions, the Lyman-$\alpha_1$ ($1s$--$2p_{3/2}$) transition in H-like ions and the K-$\alpha$ ($1s^2$--$1s2p$) transition in He-like ions represent prototypical one- and two-electron systems.
Consequently, their polarization has been extensively investigated both experimentally and theoretically \cite{Nakamura10,Robbins1,Bostock1,Reed1,Wang1,Chen12,Henderson1,Beiersdorfer7,Wu3}.
In contrast, the polarization of the $2s$--$2p_{3/2}$ transition in Li-like ions has remained largely unexplored, with only a few exceptions reported in the literature \cite{Gau1}.
Despite its structural simplicity as a single-valence-electron system and its critical role as a resonance transition responsible for intense emission in astrophysical and laboratory plasmas, relatively few studies have been reported.

An electron beam ion trap (EBIT) \cite{Marrs1,yebisu_nakamura} serves as a powerful device for investigating highly charged ions under strictly controlled laboratory conditions.
Within an EBIT, a tightly focused electron beam both produces and confines the ions, facilitating spectroscopic studies across a broad range of charge states at precisely defined interaction energies.
Because this unidirectional beam establishes a strong symmetry axis within the system, the radiation emitted subsequent to electron-impact excitation inherently exhibits anisotropy and linear polarization.
Consequently, EBIT facilities have been extensively employed to probe polarization phenomena in the emission of highly charged ions, thereby yielding essential benchmarks for both atomic theory and plasma diagnostics.

Experimental techniques employed for polarization measurements are highly dependent on the photon energy regime.
In the visible and near-ultraviolet regions, such measurements are relatively straightforward due to the ready availability of high-quality linear polarizers and waveplates.
In the X-ray region, well-established methods, such as Bragg reflection \cite{Tsuboi1,Henderson1,Beiersdorfer6,Gall1}, photoelectric polarimeters \cite{Vink1,Costa1,Iwata1}, and Compton polarimeters \cite{Hitomi2,Shah1,Nakamura34}, have been successfully utilized to determine the polarization of radiation from highly charged ions and astrophysical sources.
In contrast, polarization measurements in the vacuum-ultraviolet (VUV) region present significant experimental challenges.
The scarcity of efficient polarizing optics and suitable retardation elements complicates the construction of reliable polarization analyzers, thereby limiting the number of experimental studies focusing on VUV polarization.

In this paper, we report measurements of the polarization of the $2s$--$2p_{3/2}$ transition in Li-like N$^{4+}$ following electron-impact excitation.
The experiment was conducted using a compact EBIT equipped with a newly developed VUV spectropolarimeter \cite{Nakamura35}.
The degree of polarization of the emitted radiation was measured across an electron collision energy range of 85~eV to 1000~eV.
These results provide experimental data regarding the polarization properties of collisionally excited emission in Li-like ions, thereby contributing to the advancement of VUV polarization diagnostics for anisotropic plasmas.

%%%%%%%%%%%%%%%%%%%%%%%%%%%%%%%%%%%%%%%%%%%%%%%%%%%%%%%%%%%%%
%%%%%%%%%%%%%%%%%%%%%%%%%%%%%%%%%%%%%%%%%%%%%%%%%%%%%%%%%%%%%
\section{\label{sec:exp}Experiment}
%%%%%%%%%%%%%%%%%%%%%%%%%%%%%%%%%%%%%%%%%%%%%%%%%%%%%%%%%%%%%
%%%%%%%%%%%%%%%%%%%%%%%%%%%%%%%%%%%%%%%%%%%%%%%%%%%%%%%%%%%%%

%%%%%%%%%%%%%FIGURE%%%%%%%%%%%%%%%%%%%%%%%%%%%%%%%%%%
%\begin{figure}[tb]
%\includegraphics[width=0.48\textwidth]{setup.eps}
%\caption{\label{fig:setup}
%Schematic diagram of the experimental apparatus.
%CoBIT: compact electron beam ion trap; WP: waveplate; PA: polarization }
%\end{figure}
%%%%%%%%%%%%%FIGURE%%%%%%%%%%%%%%%%%%%%%%%%%%%%%%%%%%

The present study was performed with a recently developed spectropolarimeter \cite{Nakamura35} coupled to a compact EBIT known as CoBIT \cite{cobit}.
CoBIT consists primarily of an electron gun, a drift tube (DT), an electron collector, and a superconducting magnet featuring high-critical-temperature superconducting wire that operates at liquid-nitrogen temperature.
The DT comprises three successive cylindrical electrodes that form a potential well, enabling the axial trapping of ions by applying a higher potential to both end electrodes relative to the middle one.
An electron beam emitted from the gun is accelerated and injected into the DT, where it is compressed by the axial magnetic field generated by the magnet.
The space-charge potential of this compressed, high-density electron beam serves to trap the ions radially.
Consequently, highly charged ions are produced within the DT through the successive electron-impact ionization of trapped ions.

To produce and trap N$^{4+}$ ions, N$_2$ gas was injected into CoBIT from a gas injector coupled to one of the side ports of CoBIT.
To examine the charge exchange contribution, measurements without injecting N$_2$ gas were also performed.
Due to a minor leak from the liquid N$_2$ tank within CoBIT, N$^{4+}$ ions could still be produced and observed under these conditions, albeit at a significantly reduced signal rate.
To prevent the accumulation of heavier ions, such as Ba and W evaporated from the electron-gun cathode, the trap was periodically emptied every 0.5~s.

The details of the spectropolarimeter are given in Ref.~\cite{Nakamura35}.
Briefly, it consists of a waveplate, a grating, a polarizer used as a polarization analyzer, and a position-sensitive detector (PSD), arranged in this order along the path of the incident VUV light.
The combination of the waveplate and the polarization analyzer, which were originally developed for measuring the polarization of the Lyman-$\alpha$ in the CLASP rocket experiment \cite{Kano1,Watanabe19}, enables polarimetry measurements, whereas the combination of the grating and the PSD, which were originally used for a VUV spectrometer for CoBIT \cite{Nakamura31}, enables wavelength dispersive spectroscopic measurements.

Since the spectropolarimeter was coupled at 90$^\circ$ with respect to the electron beam, which defines the quantization axis, we can focus on the linear polarization $P$ defined as:
\begin{equation}
P =
\frac{I_\parallel - I_\perp}
     {I_\parallel + I_\perp},
\end{equation}
where $I_\parallel$ and $I_\perp$ represent the intensities of radiation with polarization vectors parallel and perpendicular to the quantization axis, respectively.
Assuming the waveplate is an ideal half-waveplate that rotates the polarization vector of the linearly polarized radiation and the polarization analyzer is an ideal polarizer, the radiation intensity detected at the PSD varies as a function of the waveplate rotation angle, and this modulation amplitude directly gives the degree of linear polarization.

The waveplate \cite{Ishikawa1} is formed by stacking two $\text{MgF}_2$ plates with slightly different thicknesses, oriented such that their principal axes are orthogonal.
The retardation $\delta$ at wavelength $\lambda$ is dictated by the thickness difference $\Delta d$ and the birefringence $n_e - n_o$ at $\lambda$ according to the relation
\begin{equation}
    \delta = \frac{2 \pi (n_e-n_o) \Delta d}{\lambda}.
\end{equation}
The waveplate functions as a half-waveplate when $\delta=180^\circ$.
In the present study, plates with $\Delta d = 8.420 \pm 0.050~\mu\text{m}$ were employed, which serves as a half-waveplate to a good approximation at the wavelength of interest (123.88~nm) \cite{Nakamura35}.
The waveplate was mounted on an ultrahigh-vacuum rotation stage (\mbox{SmarAct} SR-5714C) to observe intensity modulation as a function of the rotation angle of the waveplate.
Since several hours of data acquisition were required to obtain sufficient statistics, the waveplate rotation angle was randomly changed every 1~min to reduce the influence of long-term instrumental drifts by randomizing the measurement sequence.

The polarization analyzer \cite{Narukage1} is a fused silica plate coated with SiO$_2$ and MgF$_2$ layers, designed to operate as a reflective polarizer that predominantly reflects $s$-polarized radiation for the Lyman-$\alpha$ wavelength at a Brewster angle of approximately $68^\circ$.
The polarizing power $\mathcal{P}$ of the analyzer is defined as $\mathcal{P}=(R_s-R_p)/(R_s+R_p)$, where $R_s$ and $R_p$ represent the reflectivities for $s$- and $p$-polarized radiation.
According to the measurements by Goto et al. \cite{Goto1,Nakamura35}, $\mathcal{P}=0.99$  at the wavelength of interest (123.88~nm), i.e., the polarization analyzer functions as a reflective polarizer with an extinction ratio of about 200.
In the current configuration, the designed incident angle is $\theta = 67.5^\circ$.

The grating (Hitachi High-Tech Corporation, model 001-0639), which provides the wavelength dispersion, is an aberration-corrected, variable-line-spacing type designed to enable flat-field focusing.
It features a groove density of 600~mm$^{-1}$ and operates at an incidence angle $\alpha$ of 85.3$^\circ$.
For the $2s$--$2p_{3/2}$ transition in N$^{4+}$ at 123.88~nm, the corresponding exit angle $\beta$ is approximately 67.3$^\circ$.
The PSD incorporates five microchannel plates (MCPs) coupled with a resistive anode.
To enhance the detection efficiency for VUV photons, the front surface of the first-stage MCP is coated with CsI.

%To measure the polarization of radiation in the VUV regime, a recently developed grazing-incidence flat-field spectrometer \cite{Nakamura31} was adapted into a spectropolarimeter \cite{Nakamura35}.
%In its original configuration, the spectrometer consisted solely of a grating and a position-sensitive detector (PSD).
%The grating (Hitachi High-Tech Corporation, model 001-0639) is an aberration-corrected, variable-line-spacing type designed to enable flat-field focusing.
%It features a groove density of 600~mm$^{-1}$ and operates at an incidence angle $\alpha$ of 85.3$^\circ$.
%For the $2s$--$2p_{3/2}$ transition in N$^{4+}$ at 123.88~nm, the corresponding exit angle $\beta$ is approximately 67.3$^\circ$.
%The PSD incorporates five microchannel plates (MCPs) coupled with a resistive anode.
%To enhance the detection efficiency for VUV photons, the front surface of the initial MCP is coated with CsI.

%To facilitate the polarization measurements, a waveplate \cite{Ishikawa1} was installed between CoBIT and the grating, while a polarization analyzer \cite{Narukage1} was positioned between the grating and the PSD, as depicted in Fig.~\ref{fig:setup}.
%Both the waveplate and the polarization analyzer were originally developed for measuring the polarization of the Lyman-$\alpha$ transition from the solar corona during the CLASP rocket experiment \cite{Kano1,Watanabe19}.

\section{Analysis}
The observed intensity, $I^\mathrm{obs}$, exhibits a modulation as a function of the waveplate rotation angle $\phi$, which can be expressed as \cite{Nakamura35}:
\begin{equation}
I^\mathrm{obs}(\phi) = A \left( 1 + B \cos \left[ 4 \left( \phi - \phi_0 \right) \right] \right) + C.
\label{eq:mod1}
\end{equation}
Here, $A$ represents the average intensity, $C$ denotes the background contribution, and $\phi_0$ is the angle between the principal axis of the waveplate and the quantization axis (defined by the electron beam direction in the present setup).
When the background contribution is subtracted from the measured intensity, the parameter $C$ can be omitted.
The parameter $B$ corresponds to the modulation amplitude, from which the degree of polarization is deduced.
By taking into account the retardation $\delta$ of the waveplate and the analyzing power $\mathcal{P}$ of the analyzer, the relationship between the modulation amplitude $B$ and the polarization $P$ is obtained as \cite{Nakamura35},
\begin{equation}
P = \frac{2B}{(1-\cos \delta)-(1+\cos \delta)B} \cdot \frac{1}{\mathcal{P}}.
\end{equation}
In the case of an ideal half-waveplate ($\delta = 180^\circ$) coupled with an ideal polarization analyzer ($\mathcal{P} = 1$), the modulation amplitude $B$ directly yields the polarization $P$.

Provided that $\phi_0$ is known, the modulation amplitude $B$ can be determined from the intensities measured at two specific angles, $\phi_0$ and $\phi_0 + 45^\circ$, using the following equation:
\begin{equation}
B = \frac{I^\mathrm{obs}(\phi_0)-I^\mathrm{obs}(\phi_0+45^\circ)}{I^\mathrm{obs}(\phi_0)+I^\mathrm{obs}(\phi_0+45^\circ)-2I_{\mathrm{BG}}},
\label{eq:2points}
\end{equation}
where $I_\mathrm{BG}$ represents the background contribution, which is assumed to remain constant at both angles.

%%%%%%%%%%%%%%%%%%%%%%%%%%%%%%%%%%%%%%%%%%%%%%%%%%%%%%%%%%%%%
%%%%%%%%%%%%%%%%%%%%%%%%%%%%%%%%%%%%%%%%%%%%%%%%%%%%%%%%%%%%%
\section{\label{sec:calc}Calculations}
%%%%%%%%%%%%%%%%%%%%%%%%%%%%%%%%%%%%%%%%%%%%%%%%%%%%%%%%%%%%%
%%%%%%%%%%%%%%%%%%%%%%%%%%%%%%%%%%%%%%%%%%%%%%%%%%%%%%%%%%%%%

To facilitate comparison with the experimental results, the polarization of the $2s$--$2p_{3/2}$ transition was evaluated using the Line Polarization module of the Flexible Atomic Code (FAC v.1.1.5) \cite{FAC}.
In the theoretical model, the $1s^2 nl$ and $1s 2l nl'$ configurations ($n\leq5$, $l, l'<n$) were included.
Electron-impact excitation and de-excitation processes, as well as E1, M1, E2, and M2 radiative transitions, were taken into account.
Although configurations up to $n=6$ were examined, no significant change in the results was found.
The contribution of the Breit interaction was explicitly evaluated and found to have only a minor effect on the polarization as anticipated for an element with a low atomic number like nitrogen.
We assumed a Gaussian electron energy distribution with a central energy $E_e$, a full width at half maximum (FWHM) of $E_e/100$, and an electron density of $10^{10}$~cm$^{-3}$.
However, preliminary checks confirmed that the calculated polarization is largely insensitive to variations in both the energy spread and the electron density.

%%%%%%%%%%%%%%%%%%%%%%%%%%%%%%%%%%%%%%%%%%%%%%%%%%%%%%%%%%%%%
%%%%%%%%%%%%%%%%%%%%%%%%%%%%%%%%%%%%%%%%%%%%%%%%%%%%%%%%%%%%%
\section{\label{sec:results}Results and discussion}
%%%%%%%%%%%%%%%%%%%%%%%%%%%%%%%%%%%%%%%%%%%%%%%%%%%%%%%%%%%%%
%%%%%%%%%%%%%%%%%%%%%%%%%%%%%%%%%%%%%%%%%%%%%%%%%%%%%%%%%%%%%

%%%%%%%%%%%%%FIGURE%%%%%%%%%%%%%%%%%%%%%%%%%%%%%%%%%%
\begin{figure}[tb]
\includegraphics[width=0.45\textwidth]{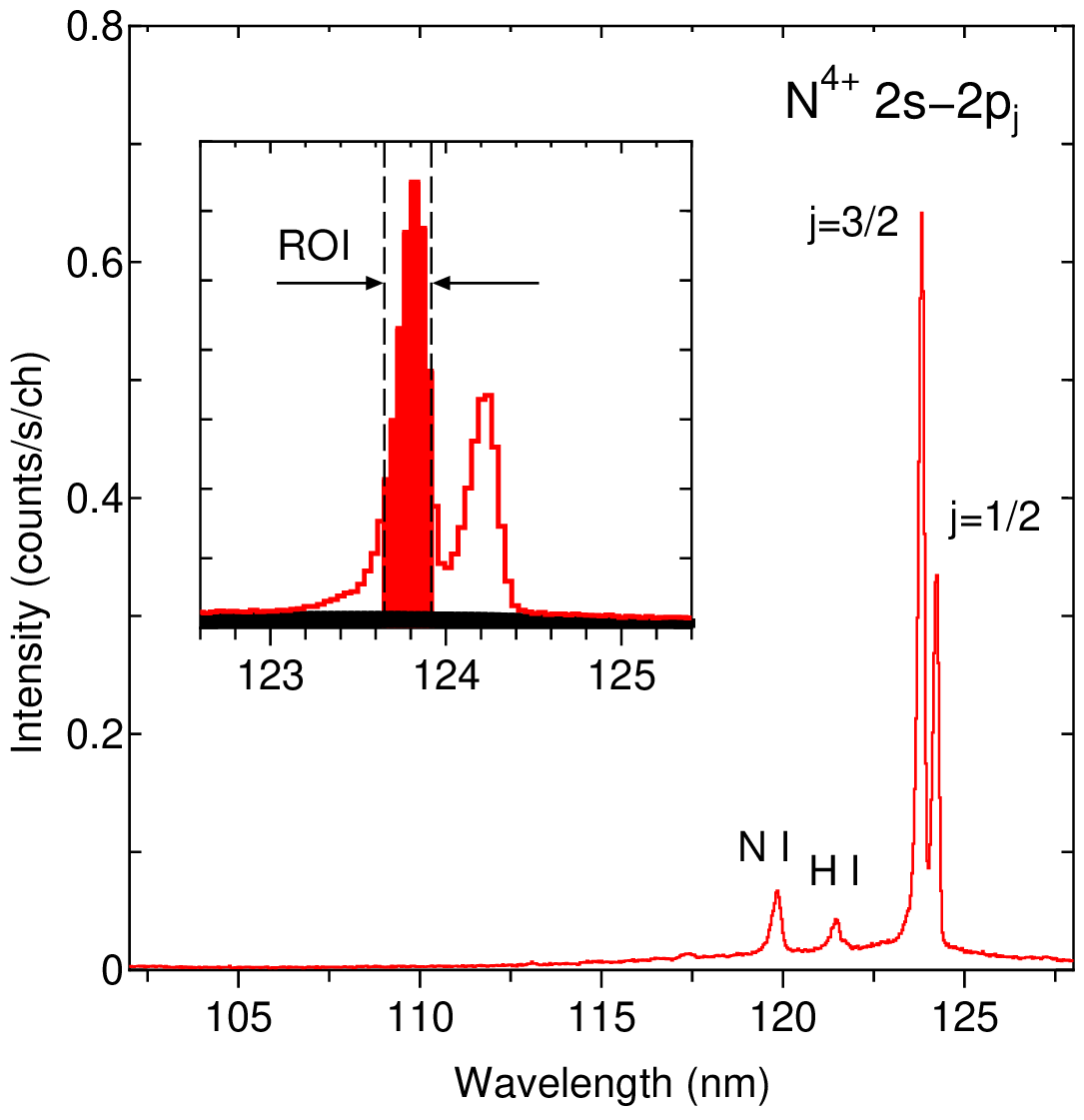}
\caption{\label{fig:spectra}
Emission spectrum measured at an electron collision energy of 1000~eV.
The inset displays an enlarged view of the $2s$--$2p$ transitions in Li-like N$^{4+}$.}
\end{figure}
%%%%%%%%%%%%%FIGURE%%%%%%%%%%%%%%%%%%%%%%%%%%%%%%%%%%

Figure~\ref{fig:spectra} displays the spectra acquired by the spectropolarimeter at an electron collision energy of 1000~eV during the injection of N$_2$ gas into CoBIT.
The $2s$--$2p_j$ ($j=1/2$ and $3/2$) transitions in N$^{4+}$ are clearly observed at approximately 124~nm.
To determine the intensity of the $2s$--$2p_{3/2}$ transition, the photon counts within the region of interest (ROI) indicated in the figure were integrated.
Background contributions were estimated from the adjacent regions on both sides of the peak (indicated in black) and subsequently subtracted from the total counts.

%%%%%%%%%%%%%FIGURE%%%%%%%%%%%%%%%%%%%%%%%%%%%%%%%%%%
\begin{figure}[tb]
\includegraphics[width=0.45\textwidth]{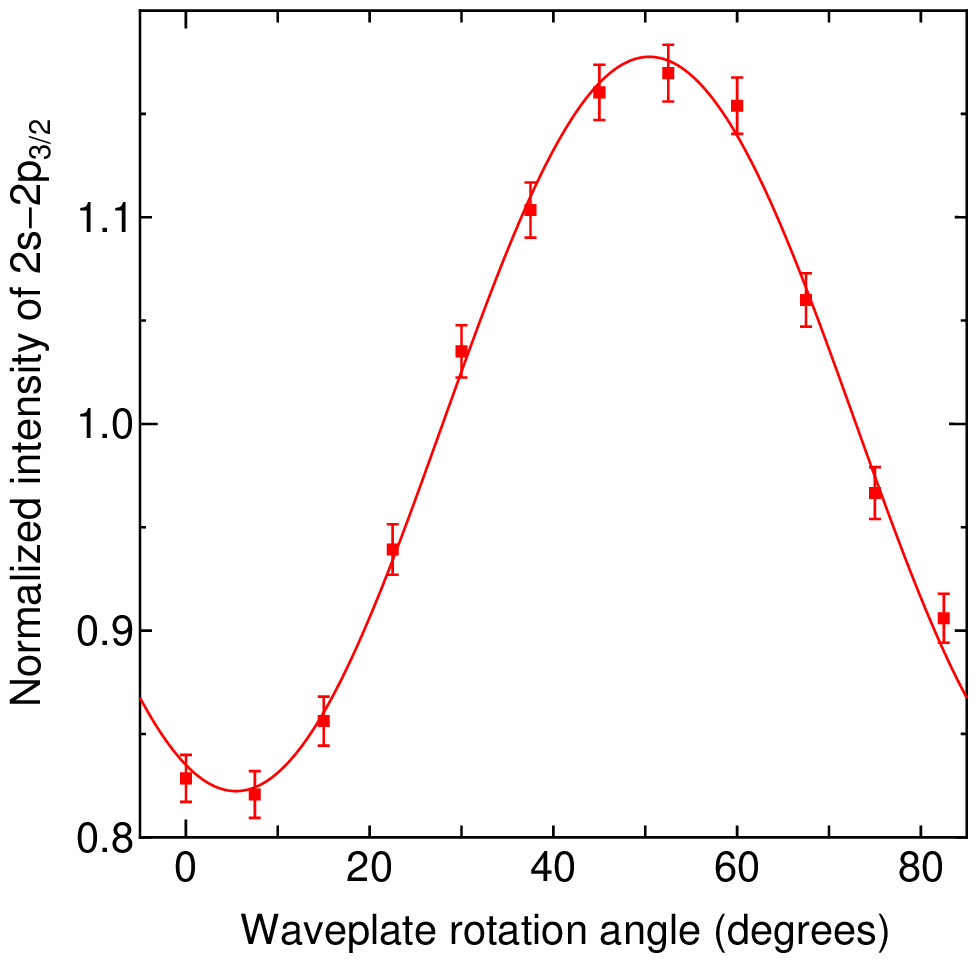}
\caption{\label{fig:mod}
Measured intensity of the $2s$--$2p_{3/2}$ transition as a function of the waveplate rotation angle.
Experimental data are shown as squares, where the error bars represent statistical uncertainties.
The solid curve illustrates a fit of Eq.~(\ref{eq:mod1}) to the data.
}
\end{figure}
%%%%%%%%%%%%%FIGURE%%%%%%%%%%%%%%%%%%%%%%%%%%%%%%%%%%

Figure~\ref{fig:mod} illustrates the intensity of the $2s$--$2p_{3/2}$ transition, measured at an electron energy of 1000~eV, as a function of the waveplate rotation angle $\phi$.
Note that $\phi=0$ is defined by the reference axis of the rotation stage, and an offset angle $\phi_0$ existed between the principal axis of the waveplate and the reference axis of the stage, although the waveplate was installed such that $\phi_0$ was nearly zero.
Because the signal acquisition time varied slightly across different angles, the time-averaged intensity is plotted here.
As depicted in the figure, a distinct modulation is evident.
The solid curve represents a fit of Eq.~(\ref{eq:mod1}) to the experimental data; the background parameter $C$ was omitted since the background contribution had already been subtracted from the intensity at each angle.
Statistical errors were employed as weights during the fitting procedure.
The best-fit values for the parameters $B$ and $\phi_0$ were determined to be $-0.178 \pm 0.004$ and $5.4 \pm 0.3^\circ$, respectively.
The modulation can also be fitted with $B=+0.178$ and $\phi_0 = 50.4^\circ$ (5.4$^\circ$+45$^\circ$).
However, as already explained, since the waveplate was installed to have $\phi_0\sim0$, $\phi_0$ should be 5.4$^\circ$ rather than 50.4$^\circ$; thus, $B$ should be negative, which means the polarization $P$ is negative, indicating that the radiation component polarized perpendicular to the quantization axis was larger.
Given that this specific dataset exhibited the most pronounced modulation and provided the $\phi_0$ value with the smallest uncertainty, this determined $\phi_0$ was adopted for all subsequent analyses.
Consequently, $\phi_0$ was held constant when extracting the modulation amplitude $B$ from the angular dependence of the intensity.
In other instances, measurements were restricted to the two specific angles $\phi = \phi_0$ and $\phi = \phi_0 + 45^\circ$, allowing $B$ to be calculated directly via Eq.~(\ref{eq:2points}).

\begin{table}[htbp]
\caption{Summary of the experimental conditions and the measured modulation amplitudes $B$.
The measured $B$ values are listed as the polarization $P$ assuming an ideal polarizer ($(R_\mathrm{s}-R_\mathrm{p})/(R_\mathrm{s}+R_\mathrm{p}) = 1$) and an ideal half-waveplate ($\delta = 180^\circ$).
Here, $\Delta P_\mathrm{sys}$ represents the systematic uncertainty arising from deviations from these assumptions, and $\Delta P_\mathrm{st}$ denotes the statistical uncertainty.}
\label{tab:results}
\begin{ruledtabular}
\begin{tabular}{ccccccc}
%\hline
$E_\mathrm{e}$ (eV) & $I_\mathrm{e}$ (mA) & method & gas & $P$ ($B$) & $\Delta P_{\mathrm{st}}$ & $\Delta P_{\mathrm{sys}}$ \\
\hline
85   & 2.2 & 0-45 & N$_2$ & 0.016  & 0.016  & 0.0008 \\
150  & 5   & 0-45 & -- & -0.014 & 0.003 & 0.0007 \\
150 & 5   & 0-45 & N$_2$ & -0.021 & 0.003 & 0.0011 \\
150 & 5  & mod & -- & -0.016 & 0.003 & 0.0008 \\
150 & 5   & mod  & N$_2$ & -0.031 & 0.004 & 0.0016 \\
280  & 13  & 0-45 & -- & -0.072 & 0.004 & 0.004 \\
280* & 13  & 0-45 & -- & -0.068 & 0.002 & 0.003 \\
280 & 14  & mod  & N$_2$ & -0.068 & 0.003 & 0.003 \\
500 & 14 & 0-45 & -- & -0.134 & 0.003 & 0.007 \\
500 & 14 & mod & -- & -0.142 & 0.005 & 0.007 \\
500 & 14 & 0-45 & N$_2$ & -0.133 & 0.003 & 0.007 \\
500 & 14 & mod & N$_2$ & -0.133 & 0.005 & 0.007 \\
1000 & 13  & 0-45 & -- & -0.185  & 0.008 & 0.009 \\
1000 & 13  & mod  & N$_2$ & -0.178  & 0.005 & 0.009 \\
%\hline
\end{tabular}
\end{ruledtabular}
\end{table}

Table~\ref{tab:results} summarizes the modulation amplitudes $B$ measured at electron collision energies of 85, 150, 280, 500, and 1000~eV.
These energies are nominal values, as they correspond simply to the potential difference between the electron gun cathode and the center of the ion trap.
Under the assumptions of an ideal half-waveplate ($\delta=180^\circ$) and an ideal polarizer ($\mathcal{P}=1$), the measured values of $B$ are presented directly as the polarization $P$ in the table.
The systematic uncertainty, $\Delta P_\mathrm{sys}$, was evaluated based on our previous study \cite{Nakamura35}.
Specifically, a maximum deviation of 10$^\circ$ in $\delta$ from 180$^\circ$ and a minimum analyzing power of 0.97 were assumed.
In addition, it was also assumed that the contribution from the unresolved $j=1/2$ peak would reduce the polarization magnitude by 0.002.
These systematic uncertainties contribute only to an increase in the absolute magnitude of $P$ relative to $B$.

To verify the robustness of our results, the polarization was evaluated under various experimental conditions.
Specifically, two measurement methods were employed.
The first method (denoted as ``mod" in Table~\ref{tab:results}) determines the modulation amplitude $B$ by fitting Eq.~(\ref{eq:mod1}) to the experimental intensity modulation, as illustrated in Fig.~\ref{fig:mod}.
The second method (denoted as ``0--45") calculates $B$ via Eq.~(\ref{eq:2points}) using the intensities measured only at two angles: $\phi=\phi_0$ and $\phi = \phi_0+45^\circ$.
For the ``mod" method, the statistical uncertainty $\Delta P_\mathrm{st}$ corresponds to the standard error of $B$ derived from the fitting procedure.
For the ``0--45" method, $\Delta P_\mathrm{st}$ was deduced through standard error propagation applied to Eq.~(\ref{eq:2points}).
Second, to assess the influence of charge-exchange processes involving the injected N$_2$ gas, additional measurements were conducted without N$_2$ injection, where N$_2$ was provided by a minor leak from the liquid N$_2$ tank within CoBIT.
Furthermore, to investigate the effect of the incident angle $\theta$ on the polarization analyzer, the data marked with an asterisk were acquired at $\theta = 69.5^\circ$ instead of $67.5^\circ$ for the other data.

%%%%%%%%%%%%%FIGURE%%%%%%%%%%%%%%%%%%%%%%%%%%%%%%%%%%
\begin{figure}[tb]
\includegraphics[width=0.45\textwidth]{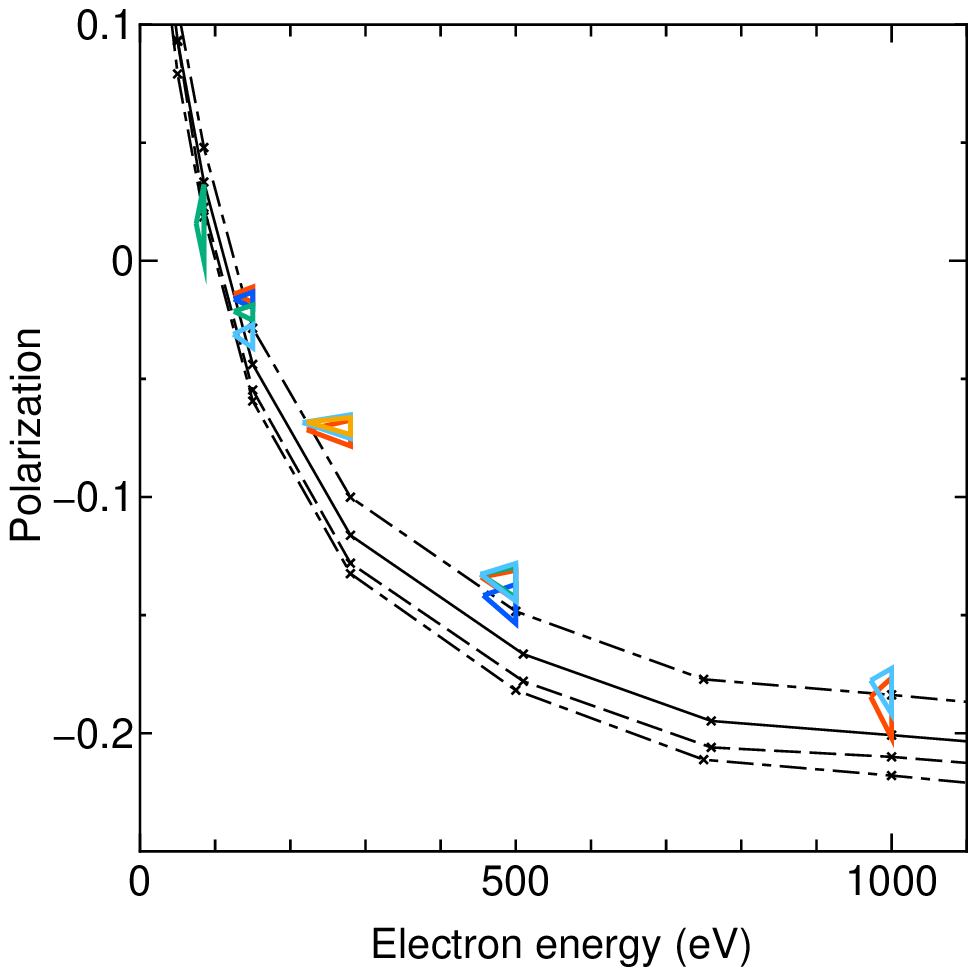}
\caption{\label{fig:results}
Measured and calculated polarization of the $2s$--$2p_{3/2}$ transition in Li-like N$^{4+}$.
The colored symbols denote the experimental results obtained in this work: red, two-angle measurements without gas injection; blue, modulation measurements without gas injection; green, two-angle measurements with gas injection; light blue, modulation measurements with gas injection; and orange, measurements at a different incident angle $\theta$.
Theoretical calculations obtained using FAC are represented by the black crosses and the solid curve.
The dashed and dot-dashed curves are described in the text.}
\end{figure}
%%%%%%%%%%%%%FIGURE%%%%%%%%%%%%%%%%%%%%%%%%%%%%%%%%%%

Figure~\ref{fig:results} plots all the measured results listed in Table~\ref{tab:results} as colored triangles alongside the theoretical calculations obtained using FAC.
The electron energy values denoted in the figure (as well as in Table~\ref{tab:results}) represent the nominal collision energies, derived simply from the potential difference between the electron gun cathode and the central DT electrode.
However, the actual electron energy experienced by the trapped ions is modified by the space charge potential generated by both the electron beam and the trapped ions, as well as the penetration of the end-cap DT potentials.
The space charge potential was estimated based on the electron beam radius predicted by Herrmann's theory \cite{Herrmann1}, assuming a 30\% compensation of this potential by the trapped ions.
Furthermore, finite element analysis of the trap configuration indicates that a 50 V potential applied to the end DT electrodes enhances the central trap potential by approximately 10 V.
The horizontal extent of each triangle represents the sum of these energy shift contributions.
The vertical extent indicates the linear sum of the statistical and systematic uncertainties, where, as previously noted, the systematic uncertainty contributes only to an increase in the absolute magnitude of the polarization.

The robustness of the measured polarization at 280, 500, and 1000~eV is supported by the consistency of the results obtained across various experimental conditions.
At 150~eV, however, the four data sets exhibit discrepancies that exceed their uncertainties.
We attribute this deviation primarily to the strong energy dependence of the polarization within this energy region.
These discrepancies can be reasonably explained by slight variations of a few eV in the actual electron energy among the different data sets.
Such energy shifts could arise, for instance, from fluctuations in the number of trapped ions, which subsequently alter the degree of compensation for the electron space charge potential.
Specifically, the trapped ion population likely differs between measurements taken with and without N$_2$ injection.
Finally, due to the exceptionally low count rate at 85~eV, only a single dataset could be acquired.

The solid line in Fig.~\ref{fig:results} shows the results of the FAC calculations.
The dashed line represents the calculated polarization when considering only the direct excitation from the ground state to the $2p_{3/2}$ level.
The small difference between the two curves indicates that the cascading contribution to the polarization is minor.
As shown in Fig.~\ref{fig:results}, the theoretical calculations reproduce the overall trend of decreasing polarization with increasing electron energy; however, they systematically overestimate the absolute magnitude of the polarization by a small amount.
While several experimental factors could potentially cause depolarization, their individual contributions are estimated to be negligible, as detailed in the subsequent discussion.

%\subsection{Line blending}
A straightforward explanation for the discrepancy with the theoretical calculations is line blending.
However, according to the NIST Atomic Spectra Database \cite{NIST_2024}, there are no other nitrogen lines near this wavelength.
While the $1s^2 2s3s\; ^1\!S_0$--$1s^2 2s5p\; ^1\!P_1$ transition in N$^{3+}$ appears at a nearby wavelength in the second-order reflection (61.9663~nm in the first order and 123.933~nm in the second order), this transition occurs between highly excited states, which are unlikely to be significantly populated in an EBIT.
Other potential contaminants include residual gases such as carbon and oxygen, as well as barium and tungsten evaporated from the electron gun cathode.
Several transitions in barium and tungsten have nearby wavelengths, for instance, the 123.8498~nm line of W$^{2+}$ and the 61.8114~nm line (123.623~nm in the second order) of Ba$^{3+}$.
Nevertheless, these possibilities can be reasonably ruled out, as no other lines from these ions were detected despite dozens of expected transitions from these ions being present within the presently observable wavelength range shown in Fig.~\ref{fig:spectra}.

%\subsection{Spiral motion of electrons}
Typically, the electron beam in an EBIT does not propagate perfectly along the longitudinal axis; rather, the electrons undergo spiral motion around the longitudinal axis.
The transverse component of the electron velocity vector inherent in this motion introduces a depolarization effect \cite{Beiersdorfer6,Takacs1,Numadate1}.
By assuming the conservation of the product of the beam area and the transverse energy, and applying Herrmann's theory of beam compression \cite{Herrmann1} with a cathode temperature of 0.1 eV, we estimate the transverse energy component and the corresponding tilt angle to be 3.1 eV and 3.2$^\circ$, respectively, at a nominal collision energy of 1000 eV.
Even assuming a larger tilt angle of 5$^\circ$, the resulting depolarization would remain below 1\%.
This depolarization effect is therefore negligible and cannot account for the discrepancy observed between our measurements and the theoretical predictions.

%\subsection{Ionization of metastable Be-like N$^{4+}$}
Be-like ions feature long-lived metastable states, such as the $^3$P term, which can build up a substantial population within an EBIT.
The inner-shell ionization of these metastable levels could, in principle, contribute to populating the $2p_{3/2}$ upper state of the Li-like ions investigated in this study:
\begin{equation}
    \mathrm{e} \;\;+\;\;\mathrm{N}^{3+*} (1s^2 2s 2p) \rightarrow \mathrm{N}^{4+} (1s^2 2p)\;\;+\;\;2\mathrm{e} .
\end{equation}
Because this specific process is excluded from our current theoretical model, it represents a potential source of the discrepancy between the experimental and calculated values.
Nevertheless, based on collisional-radiative modeling performed with FAC, the relative population of these metastable states is estimated to be only on the order of $10^{-3}$ compared to the ground state, assuming a typical CoBIT electron density of $10^{10}$ cm$^{-3}$ and a collision energy of 1000 eV.
Consequently, the depolarizing contribution from metastable Be-like ions can be considered negligible.

%\subsection{Recombination and charge exchange of He-like N$^{5+}$}
The ionization energy of Li-like N$^{4+}$ is 98~eV, meaning that He-like N$^{5+}$ ions can be produced and trapped at the investigated collision energies except 85~eV.
As a result, both radiative recombination (RR) and charge exchange (CX) involving residual or injected N$_2$ gas, followed by radiative cascades, could theoretically contribute to the population of the $2p$ states in Li-like N$^{4+}$:
\begin{equation}
    \mathrm{e} \;\;+\;\;\mathrm{N}^{5+} (1s^2) \rightarrow \mathrm{N}^{4+} (1s^2 nl) + h\nu , 
\end{equation}
\begin{equation}
    \mathrm{N}^{5+} (1s^2) \;\;+\;\;\mathrm{B} \rightarrow \mathrm{N}^{4+} (1s^2 nl) + \mathrm{B}^{+} , 
\end{equation}
where B denotes the residual or injected molecules.
Nevertheless, because the RR cross section ($\sim 10^{-25}$ cm²) is orders of magnitude smaller than the collisional excitation cross section for Li-like N$^{4+}$ ($\sim 10^{-18}$ cm²), its depolarizing contribution is effectively negligible.
While the precise magnitude of the CX contribution is difficult to estimate due to uncertainties in the target gas pressure, it can also be neglected.
This conclusion that the CX contribution is negligible is supported by the consistent polarizations observed in measurements taken with and without N$_2$ injection.

Consequently, we suggest that the discrepancy in polarization is unlikely to be explained by experimental factors, and that the discrepancy likely originates from an overestimation in the theoretical calculations.
The polarization of the $1s$--$2p_{3/2}$ transition is determined by the populations of the magnetic sublevels of the $2p_{3/2}$ level.
The dot-dashed line in Fig.~\ref{fig:results} indicates the polarization calculated by assuming deviations of only about $\pm0.01$ in the fractional populations of the magnetic sublevels from the calculated values.
This comparison demonstrates that a difference of only 0.01 in fractional populations is sufficient to bring the calculations into reasonable agreement with the experimental results, which indicates that the present measurements provide a stringent test of the calculations.
In general, accurate measurements of absolute excitation cross sections are not straightforward and typically involve uncertainties on the order of 10~\%, making it difficult to provide a stringent test of theoretical models.
In contrast, the present study demonstrates that precise measurements of polarization enable a sensitive test of the theoretical description of magnetic sublevel populations produced by electron-impact excitation.

Similar overestimations of the polarization magnitude have been reported in the literature for the Lyman-$\alpha_1$ transition in H-like Ti$^{21+}$ \cite{Nakamura10}, Ar$^{17+}$, and Fe$^{25+}$ \cite{Robbins1}.
Bostock et al. \cite{Bostock1} demonstrated that such discrepancies in highly charged ions can be explained by incorporating the Breit interaction into the models.
Wu et al. \cite{Wu3} also pointed out the importance of the Breit interaction, especially for high-energy collisions of electrons with heavy H-like ions.
However, in the present study with a low-$Z$ Li-like ion, the FAC calculations confirmed that the Breit interaction is unlikely to account for the observed discrepancy.
Further investigations, such as systematic measurements along the Li-like isoelectronic sequence, are required to fully elucidate the origin of this discrepancy between theory and experiment.

%%%%%%%%%%%%%%%%%%%%%%%%%%%%%%%%%%%%%%%%%%%%%%%%%%%%%%%%%%%%%
%%%%%%%%%%%%%%%%%%%%%%%%%%%%%%%%%%%%%%%%%%%%%%%%%%%%%%%%%%%%%
%\section{\label{sec:discussion}Discussion}
%%%%%%%%%%%%%%%%%%%%%%%%%%%%%%%%%%%%%%%%%%%%%%%%%%%%%%%%%%%%%
%%%%%%%%%%%%%%%%%%%%%%%%%%%%%%%%%%%%%%%%%%%%%%%%%%%%%%%%%%%%%

%%%%%%%%%%%%%%%%%%%%%%%%%%%%%%%%%%%%%%%%%%%%%%%%%%%%%%%%%%%%%
%%%%%%%%%%%%%%%%%%%%%%%%%%%%%%%%%%%%%%%%%%%%%%%%%%%%%%%%%%%%%
%\section{\label{sec:conclusion}Conclusion}
%%%%%%%%%%%%%%%%%%%%%%%%%%%%%%%%%%%%%%%%%%%%%%%%%%%%%%%%%%%%%
%%%%%%%%%%%%%%%%%%%%%%%%%%%%%%%%%%%%%%%%%%%%%%%%%%%%%%%%%%%%%

%\vspace{-3mm}
\begin{acknowledgments}
This work was supported by JSPS KAKENHI Grant Numbers JP24H00200 and the UEC-NAOJ matching fund project.
%XMT was supported by the JSPS KAKENHI (Grant-in-Aid for Scientific Research (C) JP22K03493.
\end{acknowledgments}

%%%%%%%%%%%%%%%%%%%%%%%%%%%%%%%%%%%%%%%%%%%%%%%%%%%%%%%%%%%%%
%%%%%%%%%%%%%%%%%%%%%%%%%%%%%%%%%%%%%%%%%%%%%%%%%%%%%%%%%%%%%
%\bibliography{ref}% Produces the bibliography via BibTeX.
%%%%%%%%%%%%%%%%%%%%%%%%%%%%%%%%%%%%%%%%%%%%%%%%%%%%%%%%%%%%%
%%%%%%%%%%%%%%%%%%%%%%%%%%%%%%%%%%%%%%%%%%%%%%%%%%%%%%%%%%%%%


\end{document}